\documentclass[12pt]{iopart}
\usepackage{amssymb,graphicx,amsbsy,url}
\begin{document}

\title{Neutral theory of chemical reaction networks}
\author{Sang Hoon Lee$^1$, Sebastian Bernhardsson$^2$,
Petter Holme$^{1,3}$, Beom Jun Kim$^4$, and Petter Minnhagen$^1$
}
\address{$^1$IceLab, Department of Physics, Ume{\aa} University, 901 87 Ume{\aa}, Sweden}
\address{$^2$FOI, Swedish Defence Research Agency, Tumba SE-14725, Sweden}
\address{$^3$Department of Energy Science, Sungkyunkwan University, Suwon
440-746, Korea}
\address{$^4$Department of Physics, Sungkyunkwan University, Suwon 440-746,
Korea}
\ead{Petter.Minnhagen@physics.umu.se}

\begin{abstract}
To what extent do the characteristic features of a chemical reaction network
reflect its purpose and function? In general, one argues that correlations
between specific features and specific functions are the key to
understanding a complex structure. However, specific features may sometimes be
neutral and uncorrelated with any system-specific purpose, function or
causal chain. Such neutral features are caused by chance and randomness.
Here we compare two classes of chemical networks: one that has
been subjected to biological evolution (the chemical reaction network of
metabolism in living cells) and one that has not (the
atmospheric planetary chemical reaction networks). Their degree
distributions are shown to share the very same neutral system-independent features.
The shape of the broad distributions is to a large extent controlled by a single parameter, the network
size. From this perspective, there is little difference between
atmospheric and metabolic networks; they are just different sizes of the same
random assembling network. In other words, the shape of the degree distribution is a
neutral characteristic feature and has
no functional or evolutionary implications in itself; it is not a matter of
life and death.
\end{abstract}


\maketitle

\section{Introduction}

The difficulty in distinguishing between design and randomness has a long
history. An early example is the watchmaker analogy by William Paley in
1802. He argues that you can conclude that a watch is designed by
studying the interconnections between its parts. From this perspective, a
living organism has been shaped into its complex form by biological
evolution, whereas an inanimate system may display a more random
complexity~\cite{wagner}.
However, some complex features in a living organism may also be neutral and essentially uncorrelated with any system-specific purpose, function or causal chain. Such neutral complex features may sometimes be difficult to identify
and have occasionally instead been attributed to
system-specific causes.

Here we compare two classes of
chemical networks: one that has been subject to biological
evolution (the chemical reaction network of metabolism in living
cells)~\cite{kegg} and one that has not (the atmospheric planetary chemical
reaction networks)~\cite{yung,Holme2011}. We show that the shapes of the degree distributions of the chemical reaction networks are just such neutral features. In fact, we show that the metabolic networks and the atmospheric networks in this respect share the very same neutral features.

Theories for neutral features, in the sense that the specific history of the system has little influence on the emerging feature, have been invoked earlier in other contexts. Two, which are akin to the ones presented here, are Hubbell's
neutral theory of biodiversity~\cite{hubbel,az} and Hatre's maximum entropy theory of ecology~\cite{harte}. These theories seek to explain species abundance distributions without invoking any knowledge of the ecological interactions and environmental factors which make up the actual causal chain. They share the basic notion with ours in that characteristic features, which apparently are outcomes of specific complex ecological processes over ages, can sometimes nevertheless be global emergent properties forced by general non-specific neutral factors. As a somewhat trivial example of a neutral feature, consider the fact that the height of a Swedish male person is undoubtedly caused be his genes and living conditions. However, the shape of the height distribution of a large collection of Swedish males at a given age is a neutral feature given by the ubiquitous Gaussian distribution.

  In this work, we discuss two neutral theories: one is the IKEA assembling network~\cite{bw,sp} and the other is the random group formation (RGF)~\cite{zip}. We clarify how they are connected and show that both give very good predictions for the various shapes of the degree distributions for chemical reaction networks. In section~\ref{sec:assemble_IKEA}, we recapitulate the basic features of the IKEA assembling network. This neutral theory predicts the shapes of the distributions using the total numbers of nodes and links, together with the number of nodes with the smallest number of links (or the smallest degree), as the sole input knowledge. Direct comparisons with the data show that the IKEA assembling network gives very good predictions of the real distributions, for both the atmospheric and the metabolic networks. As a further test of the IKEA assembling, the predicted values of the nodes with the highest degrees are compared with the data. In section~\ref{sec:RGF}, we recapitulate the basic features of the RGF theory which is a more general neutral theory: it takes into account additional, {\em a priori} unknown, relevant information. This theory instead predicts the shape of the distribution based solely on the knowledge of the total number of nodes and links together with the number of links on the node with the largest degree. The RGF version also gives a good prediction of the data. In section~\ref{sec:IKEA_RGF}, we discuss how the IKEA assembling network is related to the RGF theory. In particular, the role of the network constraint in forming the shape of the distribution is clarified. From this we argue that the shape of the distribution is an emergent property, meaning that the shape to a large extent does not depend on the explicit unknown complicated evolution chain from which it has sprung. In addition, in section~\ref{sec:atm_vs_metabolic}, we show that apart from being described by the same emergent neutral properties, the chemical atmospheric and metabolic networks have also other features in common: they can roughly be parameterized by just a single parameter, i.e., the number of connections. In section~\ref{sec:beyond_IKEA}, we discuss a feature not accounted for by the IKEA network. Finally, we present the summary in section~\ref{sec:summary}.

\section{Assembling and the IKEA network}
\label{sec:assemble_IKEA}

Chemical reaction networks are complex systems, and in particular, metabolic networks have evolved into extraordinary complex fine-tuned systems necessary for maintaining the life of an organism. This complexity is somehow reflected, to a larger or smaller extent, in any representation we choose to describe the system with. In the present case, we choose to represent a chemical reaction network as an undirected substrate-product network~\cite{Holme2010}. In this case, a substrate substance and a product substance of reactions are linked, provided both are connected through an
enzymatic reaction; in other words, a substance is a node in a linked
network. If the network has $N$ nodes and $M/2$ links, a substance on
average connects to $\langle k \rangle = M/N$ other substances. Not all
substances are connected to an equal number of others, and this difference
in connections is reflected in the distribution $P_k = N_k/N$ of degree (the number of links a node has) $k$,
where $N_k$ is the number of nodes that have $k$ connections. The question is then: what factors determine the distribution $P_k$?

The IKEA assembling network presumes that the network has been assembled in a similar way to a piece of IKEA furniture: the correct assembling is achieved by putting each piece in the correct joint, and furthermore in the correct slot of this particular joint, as well as in the correct time order. Suppose that a piece A should go to a joint with $k$ slots. Then the assembling instruction tells you which of the $N_k$ possible joints it should go to as well as which of the slots of this particular joint and in which time order. This corresponds to a total information of $\log_2(k^2 N_k)$ bits. This assembling information is the crucial characteristic of the IKEA assembling~\cite{bw,sp}. The IKEA assembling for a network is obtained by identifying the joints as the nodes and the link ends as the things which should be joined at the nodes~\cite{bw}.

\begin{figure}
\begin{center}
\includegraphics[width=0.8\textwidth]{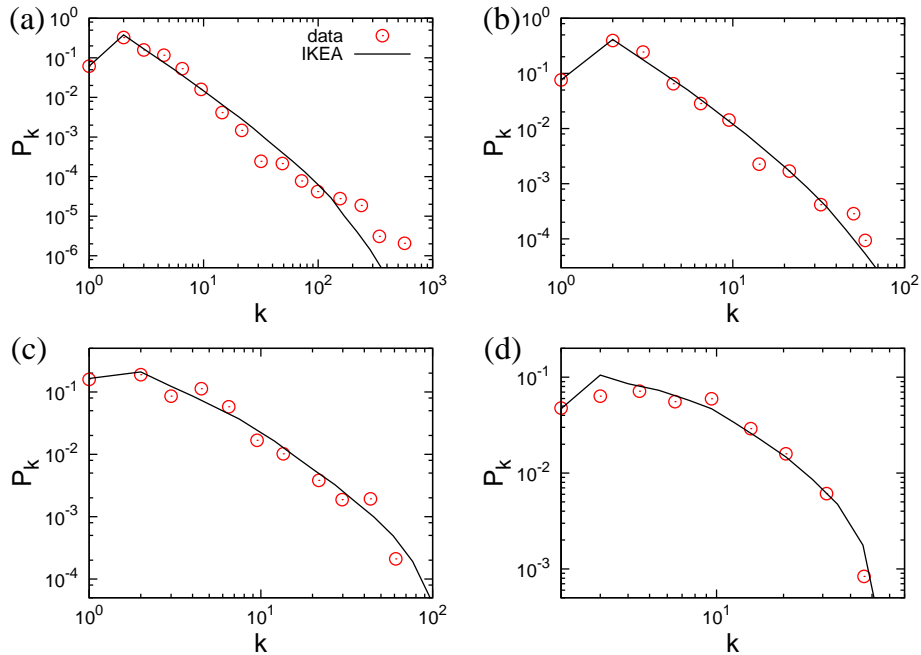}
\end{center}
\caption{Comparison of chemical reaction networks (circles) with the corresponding predictions of the IKEA assembling network (full curves). The agreement is very good for both metabolic and atmospheric network, in particular considering that the predictions in each case are only based on knowledge of the three global numbers, $N$, $M$, and $N_{k_{{min}}}$: (a) the metabolic network of the human cell for which ($N$, $M$, $N_1$)=($2218$, $12\,820$, $136$); (b) the metabolic network of {\it Mycoplasma pneumonia} for which ($N$, $M$, $N_1$)=($369$, $1534$, $28$); (c) the atmospheric network of the Earth for which ($N$, $M$, $N_1$)=($164$, $1196$, $26$); (d) the atmospheric network of Titan for which ($N$, $M$, $N_2$)=($63$, $746$, $3$).
}
\label{fig:bw}
\end{figure}

The basis for both the IKEA  and RGF predictions is the Bayesian estimate of the most likely $P_k$ for a given \emph{a priori} knowledge of the system. This corresponds to the $P_k$ which gives the maximum entropy for given constraints. Suppose that the network has $N$ nodes and $M/2$ links and further suppose that you have no \emph{a priori} knowledge of any explicit assembling instructions. This means that your best estimate corresponds to a random assembling. After you have randomly assembled it, each link end is equally likely to be found in any slot in any time order. In this case the most likely $P_k$ corresponds to the minimum of the average of the assembling information $\langle \log_2(k^2N_k)\rangle
= \sum_k P_k \log_2 (k^2N_k)$~\cite{zip}. This is equivalent to the minimum of $I_{IKEA}[P_k]=\langle \ln (k^2P_k)\rangle$ for given $N$ and $M$. However, we also \emph{a priori} know that the network is constrained to have no multiple links between nodes and no links with both link ends on the same node. In addition, the chemical networks are chemically constrained by the fact that a relatively few substances are connected to just one or two other substances (figure~\ref{fig:bw})~\cite{bw}. The IKEA assembling for chemical networks is obtained by finding the $P_k$ which minimizes $I_{IKEA}[P_k]$ for fixed $M$ and $N$ constrained by the general network constraints and the \emph{a priori} knowledge of the number of nodes with the minimum degree ($N_{k_{{min}}}$). Consequently, IKEA predictions for chemical reaction networks are only based on the \emph{a priori} knowledge of three global numbers for each network, i.e., $N$, $M$, and $N_{k_{{min}}}$.

Figure~\ref{fig:bw} compares IKEA predictions with data for four chemical networks: the metabolic networks of the human and the bacteria {\it Mycoplasma pneumoniae}, together with the atmospheric networks of the Earth and the Saturn's moon Titan. The striking thing is the agreement between the IKEA predictions and the data in all four cases, in spite of the fact that the shapes are quite different: the data for the human can to some extent over a limited range be approximated with a power law, whereas the Titan data cannot. However, an even more crucial thing to note is that the predictions are based on very little specific information of the systems: the only specific knowledge is the three numbers ($N$, $M$, $N_{k_{{min}}}$). This basically leaves you with two options: either the agreement is accidental or the shape of the distribution for a chemical reaction network is a neutral feature. We here argue for the latter case.

\begin{figure}
\begin{center}
\includegraphics[width=0.6\textwidth]{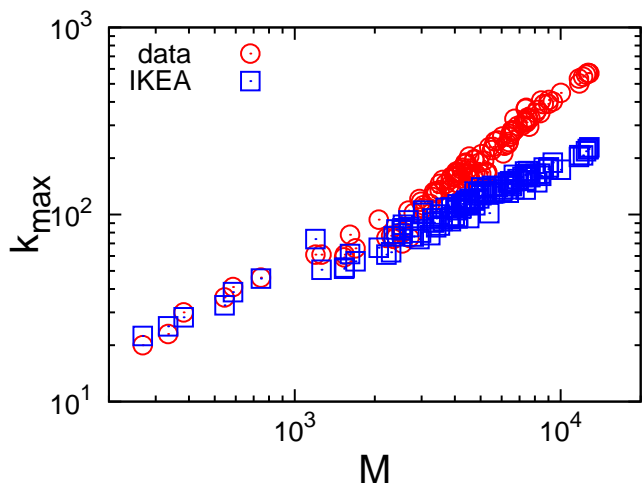}
\end{center}
\caption{Comparison between the actual number of connections to the maximum degree, $k_{max}$, and the corresponding IKEA predictions (note that $k_{max}$ is the number of connections to the most connected substance through enzymatic chemical reactions, which is water in all cases except for Titan): red circles correspond to the data. The seven data points for the smallest sizes $M$ correspond to the atmospheric data. The rest of the data correspond to metabolic networks. Note that the data approximately follow a smooth curve. The blue squares are the corresponding IKEA predictions obtained directly from the global numbers ($N$, $M$, $N_{k_{min}}$) together with the assumption of a random assembling. The agreement between the IKEA predictions and the data is striking up to network sizes of $M \simeq 3000$.}
\label{fig:kmax}
\end{figure}

Figure~\ref{fig:kmax} gives the IKEA prediction for the maximum degree, $k_{max}$. As an example, consider the chemical reaction network of Titan: the IKEA assembling predicts that, since the Titan network is a network with $M/2=373$ links, $N=63 $ nodes(= substances) and the least number of connections for a substance is $k_{{min}}=2$ of which there are in total $N_2=3$, the maximum degree is \emph{predicted} to be $k_{max} \simeq 45$. The node with the maximum degree is hydrogen for the Titan network; the IKEA prediction is that hydrogen in the Titan atmospheric network should have about $45$ connections though enzymatic chemical reactions. The actual number is $46$. If the distributions were not neutral features, you do not expect to be able to make any sensible estimate of the number of reactions involving the most connected substance based on just knowledge on the global numbers ($N$, $M$, $N_{k_{{min}}}$). As seen in figure~\ref{fig:kmax}, the IKEA predictions for $k_{max}$ are quite sensible up to network sizes of $M \simeq 3000$, regardless of whether it is a metabolic or an atmospheric network. The most connected substance is water in all cases except fot Titan. You are again left with two options: either the agreement with the IKEA predictions and the data is accidental or the distributions are neutral.

In figure~\ref{fig:kmax}, one also notes that for $M \gtrapprox 3000$ there is a systematic deviation between the IKEA predictions and the data, which grows with network size. This deviation will be discussed further in section~\ref{sec:beyond_IKEA}.

\section{Random group formation}
\label{sec:RGF}
The essential conceptual difference between the IKEA assembling and the RGF is the following: the IKEA assembling \emph{a priori} specifies \emph{all} you know about the system. The RGF, on the other hand, also takes into account knowledge which you \emph{a priori} have no explicit knowledge about, except that it is likely to exist~\cite{zip}.

The starting point of the RGF is the general formation of groups. In this case elements are collected into groups. The place in the group matters but not the time order in which the elements are assigned to the group. This means that the average information, which should be minimized in order to get the most probable $P_k$ for an RGF, is $I_{RGF}[P_k]=\langle \ln(kP_k)\rangle$. Suppose that all you know for sure \emph{a priori} is $N$ and $M$. The most probable distribution is in this case $P_k\propto \exp(-bk)/k$, where $b$ is a constant given by $N$ and $M$~\cite{zip}.

Suppose you want to apply the group formation to chemical reaction networks. Then you know that you should also consider other constraints such as the assembling constraint and the network constraints. The RGF approach at this junction argues that instead of trying to specify all these possibly missing constraints you can approximately take them into account from the fact that any additional constraint lowers the entropy of the distribution because it increases the information needed to localize a specific element. Thus you can instead lean on the assumption that the essential effect of the additional unknown constraints is a lower value of the entropy. You then obtain the most likely $P_k$ by adding the constraint that the value of the entropy is \emph{a priori} known. The form of the solution then becomes $P_k\propto \exp(-bk)/k^{\gamma}$~\cite{zip}. The actual value of the entropy is in the RGF formulation obtained by demanding that the distribution should give the correct value of the \emph{a priori} known value of the size of the largest group $k_{max}$. Thus the RGF predicts the distribution based on the \emph{a priori} knowledge of $(N,M,k_{max})$~\cite{zip}. It also has the flexibility of an arbitrary $k_{{min}}$: you can pick the subset of $N_k$ which only includes the nodes equal or larger than this new $k_{{min}}$. Provided you know the number of the remaining $N$ and connections $M$, the RGF will predict the corresponding distribution $P_{k \geq k_{{min}}}$.

\begin{figure}
\begin{center}
\includegraphics[width=0.8\textwidth]{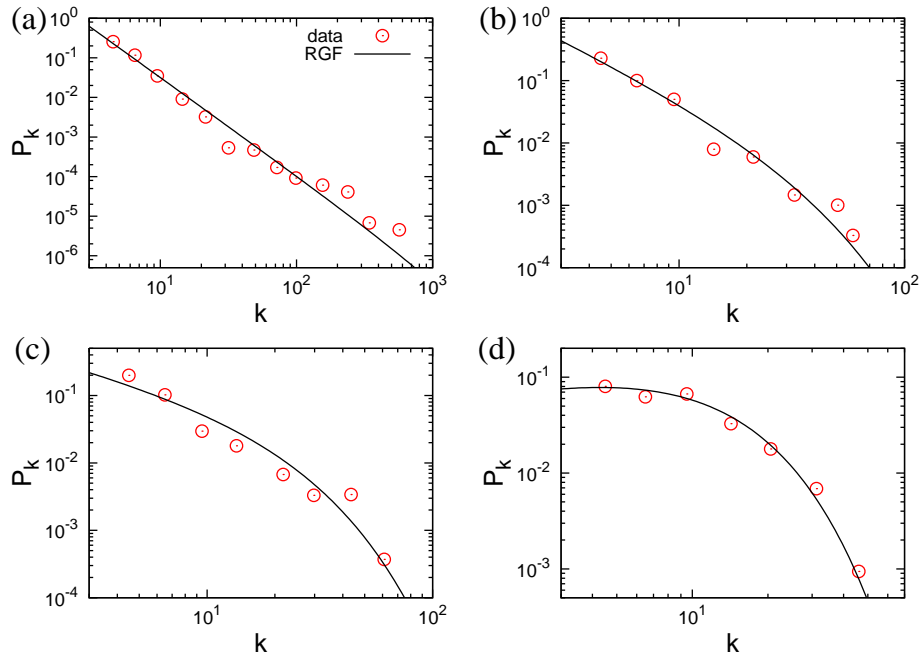}
\end{center}
\caption{Comparison between between data and RGF predictions for the same cases as in figure~\ref{fig:bw}. Red circles correspond to the data and the full drawn curves to the RGF prediction: (a), (b), (c) and (d) correspond to human, the bacteria {\em M. pneunomiae}, the Earth and the Titan, respectively. To avoid the chemical constraint for the smallest $k$ values, only the data from $k=4$ and upward are used in each case. The agreement is striking considering that the RGF prediction is based solely on the knowledge of $N$, $M$ and $k_{max}$.}
\label{fig:rgf}
\end{figure}

Figure~\ref{fig:rgf} gives the RGF predictions for the same four networks as in figure~\ref{fig:bw}. However, since the strong chemical constraint on the smallest degree is clearly outside the capability of the RGF, the analysis in each case starts from the minimum degree $k_{{min}}=4$. Note that the corresponding $M$ and $N$ values are obtained by excluding the nodes with the degree smaller than $k_{{min}}$, i.e., $M = \sum_{k=4} k N_k$ and $N = \sum_{k=4} N_k$. Again the agreement between the data and the neutral prediction provided by the RGF is striking, in particular considering that the shapes of the distributions for the metabolism of the human and the atmospheric network of Titan are quite different.

\begin{figure}
\begin{center}
\includegraphics[width=0.6\textwidth]{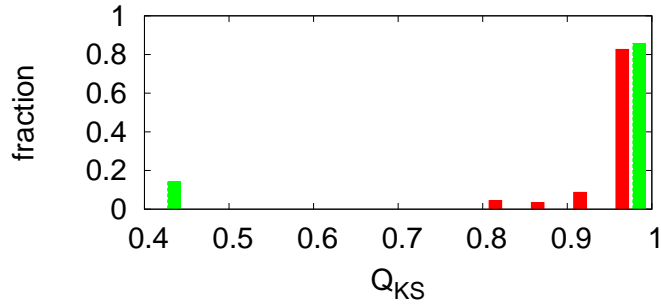}
\end{center}
\caption{Overview of the goodness of the RGF prediction for $114$ metabolic networks (red color) and $7$ atmospheric networks (green color). The horizontal axis gives the goodness in terms of the KS significance level $Q_{KS}$. The only case which does not give a very good prediction is the planet Mars ($Q_{KS} \approx 40\%$).}
\label{fig:good}
\end{figure}

Figure~\ref{fig:good} gives a measure of the goodness of the RGF predictions for $114$ metabolic networks (red) and $7$ atmospheric networks (green). According to the Kolmogorov-Smirnov (KS) test the significant level $Q_{KS}\geq 95\%$ are found for $95$ of the metabolic data-sets and for $6$ of the atmospheric (only Mars with $Q_{KS} \approx 40\%$ is singled out as different).

\section{The relation between IKEA assembling and random group formation}
\label{sec:IKEA_RGF}

\begin{figure}
\begin{center}
\includegraphics[width=0.9\textwidth]{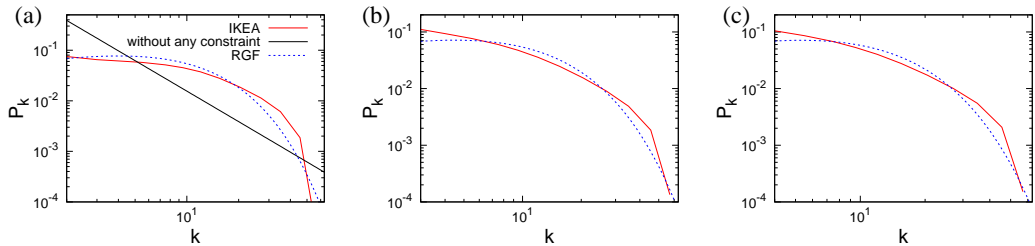}
\end{center}
\caption{RGF predictions for IKEA networks. The IKEA network for ($M$, $N$, $k_{{min}}$ =($746$, $63$, $2$)) (the same values as for Titan) is obtained and compared to the corresponding RGF predictions: (a) `straight-line-looking' curve is the unconstrained IKEA network and the full drawn curve the IKEA network including the network constraints. The dashed line is the RGF prediction; (b) IKEA network including both the network constraints and the chemical constraint $N_2=3$ (the same as for Titan); (c) the same as (b) but using data from $k=4$ and upward in the RGF prediction (instead of from $k=3$ and upward as in (a) and (b).}
\label{fig:IKEA_RGF1}
\end{figure}

In order to assess the significance of the results, we first elucidate the connection between the IKEA prediction and the RGF prediction. First suppose that your data is from an IKEA assembling with no network constraints and specified by the same numbers as for the Titan network, i.e., $M=746$, $N=63$ and the smallest degree $k_{{min}}=2$. The corresponding average $P_k$ for this unconstrained IKEA network is given by the `straight-line-looking' curve in figure~\ref{fig:IKEA_RGF1}(a). The maximum degree is on average $k_{max}=115.06$ (clearly not realizable in a real constrained network for which $k_{max} < N$) for a single realization of such an unconstrained IKEA network. Next suppose that your data is from an IKEA assembling including network constraints and again with the parameters $M$, $N$ and $k_{{min}}$ corresponding to Titan. The average $P_k$ is now instead given by the full drawn curve in figure~\ref{fig:IKEA_RGF1}(a). Thus the network constraints have a major impact on the shape of the distribution $P_k$. The largest degree is on average $k_{max}=43.5$ for a single realization of such an IKEA network, which is about half the value of the unconstrained case. The RGF prediction for this IKEA network starting from $k=3$ and upward is also given in figure~\ref{fig:IKEA_RGF1}(a) (dotted curve). The point is that the RGF takes the entropy decrease, caused by both the IKEA-assembling condition and the network constraints, into account through the value of $k_{max}$. This is of course an implicit and approximate way of taking these constraints into account. However, as demonstrated in figure~\ref{fig:IKEA_RGF1}(a), it is a very good approximation. The advantage of the RGF is that even if you do not know the true nature of the constraints you can still implicitly take them into account through $k_{max}$.

 Figure~\ref{fig:IKEA_RGF1}(b) is the IKEA network including the chemical constraint on $N_{k_{{min}}}$, which for Titan is $N_2=3$, and shows that the RGF gives a good approximation even when this additional constraint is added. Finally, for consistence, in figure~\ref{fig:IKEA_RGF1}(c), we compare the same IKEA network as in figure~\ref{fig:IKEA_RGF1}(b) with RGF, but using the data from $k=4$ and upward (instead of from $k=3$ and upward as in figure~\ref{fig:IKEA_RGF1}(b)). This shows that the starting point is not very crucial.

\begin{figure}
\begin{center}
\includegraphics[width=0.8\textwidth]{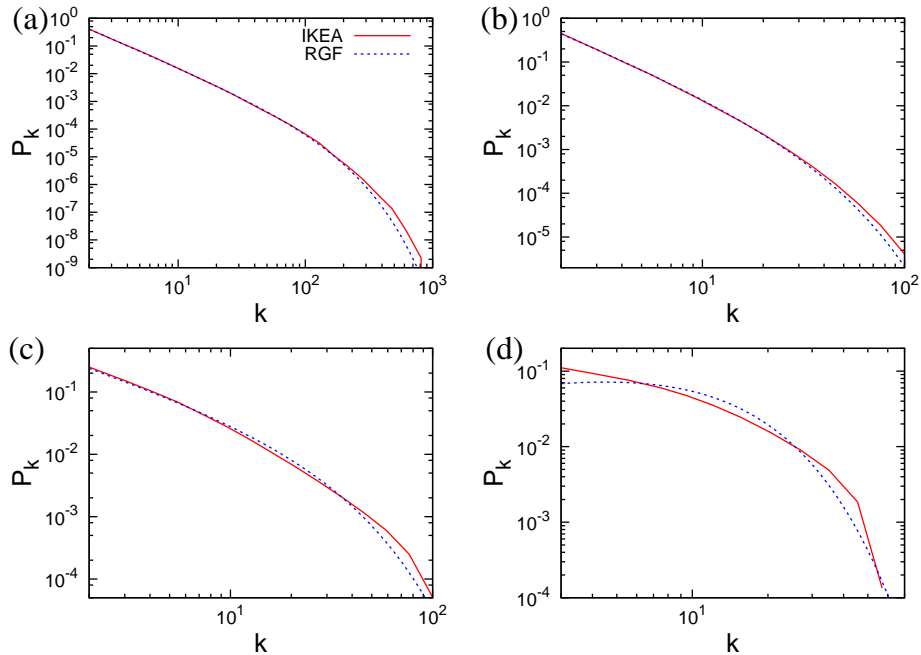}
\end{center}
\caption{RGF predictions for IKEA networks. IKEA networks for ($M$,$N$, $N_{k_{{min}}}$) including both network constraints and chemical constraints are constructed and compared to the corresponding RGF predictions. (a) ($M$, $N$, $N_{k_{{min}}}$) corresponding to the human; (b) corresponding to {\it M. pneumoniae}; (c) corresponding to the Earth; (d) corresponding to Titan. The RGF predictions are based on the data for $k=4$ and upward.}
\label{fig:IKEA_RGF2}
\end{figure}

In figure~\ref{fig:IKEA_RGF2}, we compare the IKEA networks, including both network constraints and chemical constraints with the corresponding RGF predictions for input values ($N$, $M$, $N_{k_{{min}}}$) corresponding to the human, {\it M. pneunomiae}, the Earth and Titan (compare figure~\ref{fig:bw}). The RGF is obtained for the data starting from $k=4$ and upward. The agreement is excellent in all cases. In fact, the smallest IKEA network (corresponding to Titan) shows the largest deviation from the RGF prediction. This is because for smaller networks the effect of the network constraints is larger. This means that the decrease in entropy is larger and the larger this decrease is, the less exact is the RGF.

To sum up, we have shown that the network constraints and the assembling constraints are largely sufficient for explaining the shapes of network distributions. At the same time we have explicitly shown that both these constraints, to a good approximation, can be absorbed into the RGF theory through the knowledge of the maximum degree.

\section{Comparison between atmospheric and metabolic networks}
\label{sec:atm_vs_metabolic}

So far, we have argued that the shape of the degree distribution for a chemical network is a neutral feature, because it can \emph{ipso facto} be accounted for by using very little explicit information through the random system-independent IKEA assembling network. This was further corroborated by using the RGF theory. We next show that direct comparison between the metabolic and atmospheric networks gives further evidence for neutrality.

\begin{figure}
\begin{center}
\includegraphics[width=0.6\textwidth]{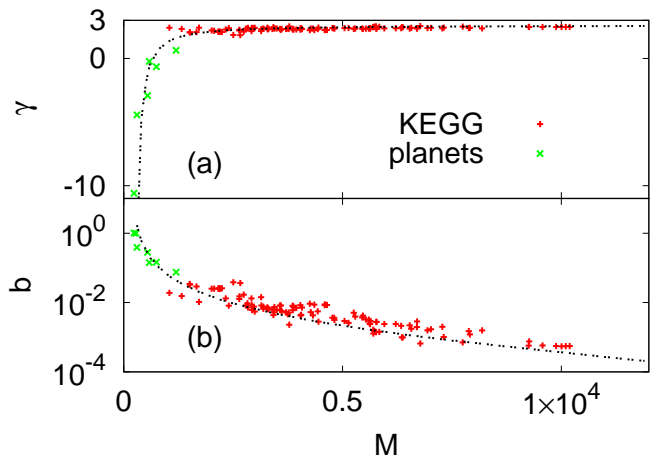}
\end{center}
\caption{RGF parameters $\gamma$ and $b$ obtained for $114$ metabolic networks (red crosses) and $7$ atmospheric networks (green crosses). The dotted curves illustrate that the data for both $\gamma$ and $b$ to a large extent fall on single curves. Note that these two parameters completely determine the shape of $P_k$. The dotted curves are predicted by RGF by assuming the linear size dependences given by the straight lines in figure~\ref{fig:linear} (see text).}
\label{fig:parameters}
\end{figure}

\begin{figure}
\begin{center}
\includegraphics[width=0.8\textwidth]{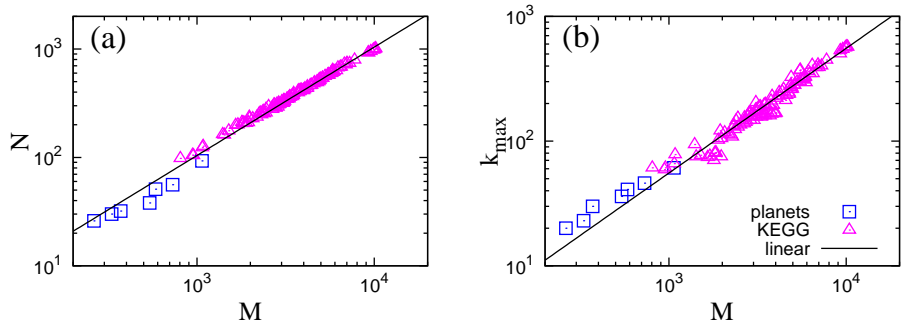}
\end{center}
\caption{$M$ dependence of $N$ and $k_{max}$. The data for $N$ and $k_{max}$ for the same $114$ metabolic networks and the $7$ atmospheric networks as in figure~\ref{fig:parameters} are plotted against $M$. The data for both $N$ and $M$ fall to a large extent on single curves. The straight lines correspond to simple linear relations $N \propto M$ and $k_{max}\propto M$.}
\label{fig:linear}
\end{figure}

Figure~\ref{fig:parameters} shows the predicted RGF parameters $\gamma$ and $b$ for the $114$ metabolic and $7$ atmospheric networks investigated here. In this analysis we used data for $k \geq 4$ (the same as figure~\ref{fig:rgf}). The point is that these parameters account for the shapes of $P_k$ for all these networks, as illustrated in figure~\ref{fig:good}. Figure~\ref{fig:parameters} shows that as a function of the total number of connections of nodes $M$ ($M= \sum_{k=4} k N_k$), both $\gamma$ and $b$ to a good approximation follow single curves. Furthermore, the atmospheric networks (green crosses in figure~\ref{fig:parameters}) and the metabolic networks (red crosses in figure~\ref{fig:parameters}) fall on the same curve. These again strongly suggest that the shapes of $P_k$ to a large extent are determined by just the size $M$ and not by the detailed origin and evolution of the network.
The dotted curves in figure~\ref{fig:parameters} are predicted from RGF by using the simple linear relationships $M\propto N$ and $k_{max} \propto M$ given by the straight lines in figure~\ref{fig:linear}. These dotted curves emphasize that the shapes of the degree distributions to a good approximation are just determined by the size $M$.

Figure~\ref{fig:linear} gives an alternative way of drawing this conclusion on the same data: the figure shows $N$ ($= \sum_{k=4}N_k$, as described in section~\ref{sec:RGF}) and $k_{max}$ as a function of $M$ ($= \sum_{k=4} k N_k$). Again the data to a good approximation follow single curves and again the same curves for atmospheric and metabolic networks. The lines in figure~\ref{fig:linear} are linear relationships suggesting that both $N$ and $k_{max}$ are roughly proportional to $M$. Figure~\ref{fig:linear}, in a very direct way, suggests a neutral origin of the degree distributions.

 From this point of view, the most crucial factor determining the shapes of the distribution is just the network sizes: the difference in the shapes of the degree distribution for the human metabolic network and Titan's atmospheric reaction network (figure~\ref{fig:bw}) is, according to this, roughly accounted for by the fact that the Titan network is about ten times smaller. The fact that one belongs to a living cell and the other to a distant atmosphere seems to have little influence on the shapes of their degree distributions.

\section{Beyond the IKEA network}
\label{sec:beyond_IKEA}

As shown in figure~\ref{fig:kmax}, the assumption of an IKEA assembling, together with the network constraints and the chemical constraint of the node with the smallest degree, is sufficient to account for the shapes of the network distributions $P_k$ for all the networks up to about $M \simeq 3000$, regardless of whether they are atmospheric or metabolic networks or not. However, the IKEA network does not give correct $k_{max}$ for larger $M$. The largest network is the metabolic network for the human, which has $M=12\,820$, whereas Titan has only  $M=746$. Comparing the IKEA prediction for these two networks shown in figure~\ref{fig:bw}(a) and (d), respectively, illustrates the point: The IKEA prediction for the human shows a deviation towards lower $P_k$ for the largest $k$.

We observe that since the IKEA network in itself contains very little system-specific information, it is really the deviations from the IKEA network that is interesting, when it comes to understanding system-specific features. The issue is, hence, what features could cause the deviation. Since the deviation is directly reflected in $k_{max}$, it is from the point of view of the RGF description signaling an additional constraint. This additional constraint is implicitly included in RGF prediction. Furthermore, since the RGF prediction works equally well for the large networks (figure~\ref{fig:good}), the implication is that the additional constraint again has a rather general character.

In \cite{sp}, it was shown that this additional constraint is also reflected as a slight lowering of the assortativity compared to the IKEA network. This means a slightly larger tendency for nodes of different degrees to connect to each other. This tendency is also reflected in a small bump for larger $k$ in the $P_k$ distribution~\cite{sp}. This bump is discernable in figure~\ref{fig:bw}(a) around $k=200$--$300$.

To sum up, the deviation in figure~\ref{fig:kmax} between the IKEA prediction and the data for the largest networks is also reflected in a slight lowering of the assortativity and the appearance of a large-$k$ bump in the distribution. This could be related to an evolutionary selection as suggested in \cite{sp}. However, the strong equivalence between the chemical reaction networks for atmospheres and metabolic networks might instead point to some additional general chemical constraints.

\section{Summary}
\label{sec:summary}

This investigation gives evidence for the existence of neutral emergent features in the context of chemical reaction networks. The conceptual basis is that some features of a complex system emerge from the complexity itself, rather than from some specific features of the system~\cite{pm}. We here have found evidence for emerging neutral features by explicitly showing that the IKEA network, which just presumes that the network has been randomly assembled, predicts the shape of the degree distribution for both atmospheric and metabolic networks to a high degree using very little information.

The IKEA network was shown to be connected to a more general neutral theory, the RGF. The RGF was shown in~\cite{zip} to describe a variety of entirely different complex systems. This suggests that neutral emergent features in complex systems are quite common and that networks are no exception in this respect.

Nevertheless, it leads to for us quite surprising results. In particular, it was shown that if you know the number of nodes, number of connections and the smallest degree, then the IKEA network theory to a good approximation predicts the number of connections water (or hydrogen in cases of Titan) has through chemical reactions, both in the case of atmospheric networks and the smaller metabolic networks. You could, of course, argue that this is just an accidental result. However, on the basis of the evidence presented, we suggest that it is a consequence stemming from the complexity itself.

One issue, which has been raised in the past, is whether or not the shape of the degree distribution in itself is signaling some evolutionary selection~\cite{wagner}. For example, it has been argued that the more power-law-like distribution for the human (see figure~\ref{fig:bw}(a)) has an evolutionary advantage over the more exponential distribution for Titan (see figure~\ref{fig:bw}(d))~\cite{wagner}. However, from the present investigation we conclude that there is no such evolutionary selection for the \emph{shape} of degree distribution: both types of distributions are equally well predicted by the same neutral IKEA network. The shape of the distribution is indeed not a matter of life and death.

\ack
This work was supported by the Swedish Research Council (SHL and PH),
the Swedish Research Council under grant no. 621-2008-4449 (PM), the
WCU program through a National Research Foundation of Korea (NRF) grant funded by
the Korean government (MEST) R31-2008-10029 (PH), and another NRF grant funded
by MEST 2011-0015731 (BJK). The authors are grateful to Andreea Munteanu and Mikael Huss for
help with data acquisition.

\section*{References}

\end{document}